\documentclass[letterpaper, 10 pt, conference]{ieeeconf}  %

\IEEEoverridecommandlockouts                              %
\expandafter\def\expandafter\normalsize\expandafter{%
	\normalsize%
	\setlength\abovedisplayskip{1pt}%
	\setlength\belowdisplayskip{2pt}%
	\setlength\abovedisplayshortskip{-8pt}%
	\setlength\belowdisplayshortskip{2pt}%
}

\usepackage{graphics} %
\usepackage{epsfig} %
\usepackage{mathptmx} %
\usepackage{times} %
\usepackage{amsmath} %
\usepackage{amssymb}  %
\usepackage[official]{eurosym}
\usepackage{graphicx}
\usepackage{relsize}
\usepackage{eso-pic}
\usepackage{xcolor}
\DeclareMathAlphabet{\mathcal}{OMS}{cmsy}{m}{n}
\SetMathAlphabet{\mathcal}{bold}{OMS}{cmsy}{b}{n}

\title{\LARGE \bf
Predictive operation of multi-energy systems in sequential markets:\\ a case study}
\author{Taouba Jouini$^{1}$, Astrid Bensmann$^{2}$, Torsten Lilge$^{1}$, Richard Hanke-Rauschenbach$^{2}$ and Matthias Müller$^{1}$  
\thanks{*This work has received funding from the German Research Foundation (DFG) – 279734922.}%
\thanks{$^{1}$ Department of Automatic Control, $^{2}$ Institute of Electric Power Systems,
        Appelstrasse 11, 30167, Hanover, Germany. 
        E-mails:
        \tt\small \{jouini, lilge, mueller\}@irt.uni-hannover.de,  \{astrid.bensmann\}@ifes.uni-hannover.de, \{hanke-rauschenbach\}@ifes.uni-hannover.de.}}%

\graphicspath{{./fig/}}

\usepackage{amsmath,amsthm,amssymb,mathrsfs,dsfont}
\usepackage{amsfonts}
\usepackage{siunitx}
\usepackage[colorinlistoftodos,prependcaption,textsize=tiny]{todonotes}
\usepackage{verbatim} 
\usepackage{tikz}
\usetikzlibrary{matrix}
\usepackage{cite}
\usepackage[colorlinks=true,
            urlcolor=blue!60!gray!90!green,
            linkcolor=blue!70!green]{hyperref}
\usepackage[normalem]{ulem}

\renewcommand{\baselinestretch}{0.957}

\newcommand\oprocendsymbol{\hbox{$\blacksquare$}}

\newcommand\oprocend{\relax\ifmmode\else\unskip\hfill\fi\oprocendsymbol}

\newcommand{\real}[0]{\mathbb R}

\newcommand{\tg}[0]{\color{green!50!black}}

\newcommand\rout{\bgroup\markoverwith{\textcolor{red}{/}}\ULon} %

\newcommand\DOItext{\color{black!70!white}\fbox{\begin{minipage}{0.815\paperwidth}
\centering\footnotesize
This is the author's version of an article that has been published in the ECC 2024 proceedings. Changes were made to this version by the publisher prior to publication.
The final version of record is available at \href{https://doi.org/10.23919/ECC64448.2024.10591115}{DOI 10.23919/ECC64448.2024.10591115}
\end{minipage}}}

\newcommand\CRtext{\color{black!70!white}\fbox{\begin{minipage}{0.815\paperwidth}
\centering\small
Copyright (c) 2024 European Control Association (EUCA). Only for non-commercial use.
\end{minipage}}}

\AddToShipoutPicture{\begin{tikzpicture}[remember picture,overlay, align=center]\node[anchor=north,yshift=-15pt] at (current page.north) {\DOItext}; \end{tikzpicture}}

\AddToShipoutPicture{\begin{tikzpicture}[remember picture,overlay, align=center]\node[anchor=south,yshift=15pt] at (current page.south) {\CRtext}; \end{tikzpicture}}

\begin{document}

\maketitle
\thispagestyle{empty}
\pagestyle{empty}

\begin{abstract} %
 We study the dispatching of multi-modal energy systems (MMES) from a sequential market perspective based on hierarchical Model Predictive Control (MPC). In a sequential setting, an upper level determines the purchased electrical power from the day-ahead market, followed by a lower-level MPC responsible for the dispatching of the multi-energy system according to the continuous trading.     
   Our case study consists in an MMES in Hanover, Germany, with electrical and heat demands as well as photovoltaic and wind energy generation, storage and coupling elements.
  We show that the hierarchical MPC solution can be embedded within the European market and German market area to provide a judicious dispatching of the MMES, also under imperfect uncertainty forecast. In particular, we discuss a reasonable choice for the prediction horizons and the effect of the forecast on the total incurring cost.

\end{abstract}

\section{Introduction}
Multi-modal energy systems (MMES) play an essential role in decarbonizing the energy system. Thereby, varying energy carriers can supply the demands in different energy forms (electricity, heat, cold)~\cite{LINGMIN2020118536,WANG2022118528}. However, with an increasing number of technical components, the operation of these systems becomes more and more complex. Additionally, the generation side fluctuates due to the integration of renewables, and the future demands are generally unknown. A market-based perspective allows for judicious management of MMES while considering economic and operational incentives, e.g., deciding when and how much power needs to be purchased from or supplied to the electrical power grid.

In sequential markets, electricity is traded at different times for a specific fulfillment date. For example, one day before delivery, an estimate of the electrical grid power over a day is purchased in the day-ahead (DA) market and a scheduling plan is determined, based on a forecast of the generation, demand and (possibly) market prices. Any deviations from the estimated grid demand can be compensated by the continuous trading (CT), within the intra-day market, while considering the physical and operational constraints. Here again, based on a more accurate forecast of demand, generation and market prices, the real-time dispatching of the MMES is resolved.

Model predictive control (MPC) is a numeric-based controller, that relies on repeated open-loop online optimization in a receding horizon fashion~\cite{rawlings2017model}. In fact, only the first input of the optimal solution is applied to the control system and the optimization problem is repeatedly solved, each time a new measurement is available, blending between open and closed-loop control. %
MPC is an engineering solution suited for multi-modal energy management because it systematically accommodates physical and operational constraints. Moreover, MPC is a computationally tractable approach thanks to the increased computational power and various efficient solvers developed in recent decades. %
Finally, forecasts about future uncertainties such as generation and demand can be effectively incorporated into the control design.

In hierarchical MPC, two optimization problems are solved sequentially on different time scales. In particular, an upper-level MPC operating at a slow time scale for the long-term decision making of storage planning provides a reference to track by a lower-level MPC operating at a faster time scale solving for the short-term. Both optimization problems account for the operational and physical constraints of the system under consideration. Due to the existing time-scale separation between DA and CT, hierarchical MPC is a well-suited approach for energy management problems.

Hierarchical MPC-based schemes prove to be an effective approach to tackle optimal operation also within sequential markets.
In~\cite{8231216}, an economic MPC is applied to Combined-Heat-Power (CHP) units based on hierarchical MPC. Therein, the upper-level MPC is a two-stage optimal control problem determining the DA schedule. The electrical grid power is forwarded to the lower-level MPC that penalizes deviations from the predicted grid power and directly controls the CHP unit in smart buildings. The authors combine heat and electricity systems together with heat market considerations and the robustness of the algorithm is tested via worst-case scenarios. In~\cite{SCHULZE2016592}, the DA and intra-day markets are considered separately with focus on stochastic optimization. The authors in~\cite{JI2021107133} consider a sequential market, while combining robust and stochastic optimization for the DA and intra-day optimization with energy and reserve scheduling under wind uncertainty and generation outages. The two-staged problem formulation consist in one deterministic, i.e., robust and one stochastic optimization layer. Robust optimization uses additive uncertainty to account for the unknown load demand and generation. A multi-energy microgrid with co-generation is considered in~\cite{WANG2022118528} together with sequential markets for both electricity and gas. In~\cite{9171315}, a hierarchical MPC has been proposed for the energy management of multiple microgrids applied to the Iberian electricity market.

 Compared to the majority of existing literature (including~\cite{9171315}) that focuses on electricity and/or heat only, we study an energy system of three forms: electricity, heat and natural gas. Each of these systems has coupling elements to another system, eventually consisting of a generation, demand and storage. We consider the sequential electricity market to decide on how and when to dispatch the MMES following the regulations of the European electricity market and German market area. For this, we leverage a hierarchical MPC scheme to solve for DA market and CT sequentially. 
Compared to~\cite{WANG2022118528}, we consider a more realistic cost in the DA market that, in addition to the electricity market costs, takes also into account the operational costs of running the multi-energy system, in particular the monetary costs for using storage devices and coupling components. This trade-off allows a favorable operation of the electrical, heat and gas systems.
Additionally, we propose an electricity market cost in the CT that penalizes the deviation with respect to the purchased grid power in the DA market to allow for monetary gain if the real-time electrical grid power is over-estimated or a penalty to pay if under-estimated in the DA market. We showcase the usefulness of the hierarchical MPC approach in dispatching the multi-energy systems in a numerical case study in Hanover, Germany, where we discuss reasonable choices of the prediction horizons and study the effect of imperfect data forecast on the total incurring cost. %

\section{Problem formulation}
\begin{figure}
	\centering
	\includegraphics[width=0.7\linewidth]{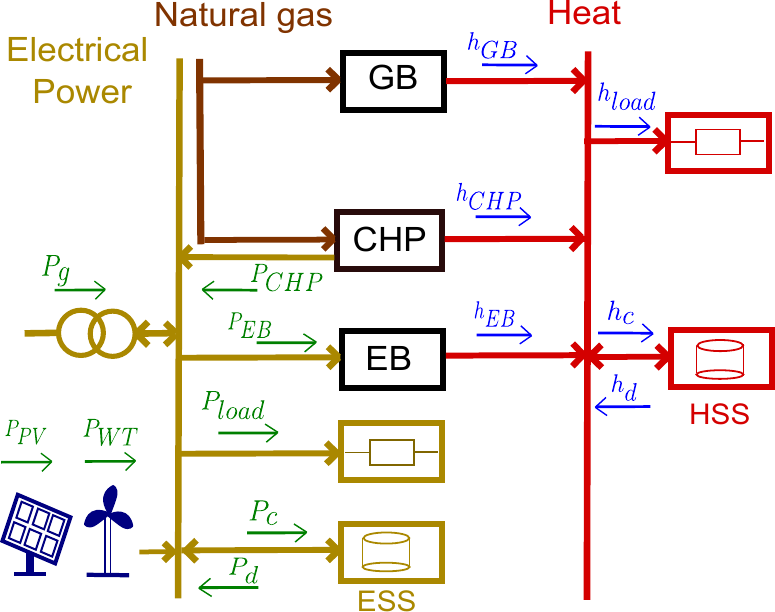}
	\caption{Overview of the multi-energy system under study with the corresponding electrical and heat power flows}
	\label{fig: multi-energy-system}
\end{figure}

We consider a multi-energy system from~\cite{WANG2022118528} with an interplay between three forms of energy, namely electricity, heat, and natural gas. Figure~\ref{fig: multi-energy-system} depicts the multi-energy system under consideration together with the associated power and heat flows. The electrical and heat system are represented each by a generation, load, storage unit, and coupled to the gas system. The coupling elements consist in the CHP plant, Electric (EB) and Gas boiler (GB).

\subsection{Physical constraints}
\label{subsec: physical-cons}
	\paragraph{Electrical system} The electrical power is generated from renewable energy resources, namely photovoltaic and wind turbines. The electrical system is connected to the power grid through the point of common coupling. Any excess of electrical power is stored into a battery representing the Electricity Storage System (ESS). 
 The State of Charge (SoC) denoted by $x_{SoC}(t)\in[0,1],\,t\in\mathbb{Z}$ follows the discrete-time system dynamics and constraints given by
\begin{subequations}
	\label{eq: const-Pc-Pd}
	\begin{align}
		x_{SoC}(t+1)&=x_{SoC}(t) + \frac{P_{c}(t)\eta_c-P_{d}(t)/\eta_d}{Q}  \delta, \label{eq: dyn-soc} \\
		\underline x &\leq x_{SoC}(t)\leq \overline x, \label{eq:soc-lim} \\
		0&\leq P_{c}(t)\leq  \overline P_{ESS}\, \alpha(t), \label{eq:charg} \\
		0&\leq P_{d}(t)\leq  \overline P_{ESS}(1-\alpha(t)), \label{eq:dis} \\
		\alpha(t)&\in\{0,1\}, 	\label{eq:bin}
	\end{align}	
\end{subequations}
where $Q>0$ is the capacity of the ESS and $\eta_c,\eta_d>~0$ are charging and discharging efficiencies, $\overline P_{ESS}>0$ is the maximal allowed charging and discharging power, $\delta$ denotes the sampling period, $\overline x$ and $\underline x$ are the maximal and minimal allowed SoC ranges. 
Note that $P_{c}(t), P_{d}(t)$ are charging and discharging powers. They constitute the inputs to the ESS and will be determined in the sequel.
In the inequalities~\eqref{eq:charg} and~\eqref{eq:dis}, $\alpha(t)$ is a binary variable that does not allow for simultaneous charging and discharging and keeps the power within an allowed range.

\paragraph{Coupling elements} 
The CHP couples natural gas to heat and electrical systems. The generated electrical power $P_{CHP}(t)$ needs to satisfy the constraints
\begin{subequations}
\label{eq: chp-eq}
	\begin{align}
		\underline P_{CHP}&\leq P_{CHP}(t) \leq \overline P_{CHP}, \\
		h_{CHP}(t) &= b\, P_{CHP}(t),
	\end{align}
\end{subequations}
where $b>0$ is a given constant characteristic of the heat-to-electricity ratio.
Here $\underline P_{CHP}>0$ and $\overline P_{CHP}>0$ represent the minimal and maximal electrical power of the CHP system.
Moreover, the EB and GB transform the electrical power and natural gas into heat, respectively. The generated heat powers $h_{EB}$ and $h_{GB}$ satisfy the following constraints
\begin{subequations}
	\begin{align}
		\underline h_{EB}&\leq h_{EB}(t)\leq \overline h_{EB},\\
		P_{EB}(t)&=\frac{h_{EB}(t)}{\eta_{EB}},\\
		\underline h_{GB} \leq & h_{GB}(t)\leq \overline h_{GB}.
	\end{align}
	\label{eq: coupling-ineq}
\end{subequations}
In~\eqref{eq: coupling-ineq}, the electrical power of the EB is denoted by $P_{EB}(t)$ and $1>\eta_{EB}>0$ is the electricity-to-heat efficiency. The upper and lower limits on heat power of the EB and GB are given by $\underline h_{EB},\,\overline h_{EB}$ and $\underline h_{GB},\,\overline h_{GB}$.

\paragraph{Heat storage system} Heat is generated through the CHP, EB and GB to satisfy heat demand. The heat is stored in a sensible Heat Storage System (HSS) and monitored at a given time $t$ via the stored heat energy $H(t)>0$. The discrete-time HSS dynamics and constraints are represented by
\begin{subequations}
\label{eq: const-hc-hd}
	\begin{align}
	H(t+1) &= H(t)+\left(h_{c}(t) \eta^h_c -\frac{h_{d}(t)}{\eta^h_d}\right) \delta, \label{eq: dyn-h}\\
		\underline{H}&\leq H(t)\leq \overline{H}, 	\label{eq: heat-bounds}\\
		0 &\leq  h_{c}(t)\leq \overline h_{HSS} \,\beta(t) \label{eq: ch-heat},\\
		0 &\leq  h_{d}(t)\leq \overline h_{HSS} (1-\beta(t)), \label{eq: dis-heat}\\
		\beta(t)& \in\{0,1\}.	\label{eq: bin-heat}
	\end{align}
\end{subequations}
In~\eqref{eq: dyn-h}, $\eta^h_c, \eta^h_d>0$ are charging and discharging efficiencies of the HSS and the upper and lower values of the stored heat energy are given by $\overline H,\underline H$ in~\eqref{eq: heat-bounds}. Moreover, $\overline h_{HSS}>0$ is the maximal heat power. Analogous to ESS, $h_c(t)$ and $h_d(t)$ represent the charging and discharging heat power and are the main input to the HSS. Here again, the binary variable $\beta(t)$ defined in~\eqref{eq: bin-heat} does not allow simultaneous charging and discharging and keeps the heat power within known limits.

\paragraph{Power balance}
Let $P_g(t)>0$ denote the electrical power fed out from (purchased) and $P_g(t)<0$ flowing into (sold) the power grid. The electrical grid power should stay within specified lower and upper limits $\underline P_g, \overline P_g$
	\begin{align}
		\underline P_g	\leq P_g(t)\leq \overline P_g\,.
		\label{eq: pow-grid-constr}
	\end{align}
	Let the electrical power generated by photovoltaic and wind turbine be given by $P_{PV}$ and $P_{WT}$, and $P_{load}$ and $h_{load}$ represent the electrical load and heat power demand modeled by an input time series.
	 In summary, the electrical and heat powers must satisfy the generation-demand balance, namely that, at each time instance, the aggregated supply of power matches the demand of electrical and heat systems, respectively
	 \begin{subequations}
	 	\begin{align}
	 	P_g+P_{PV}+P_{WT}+P_{d}+P_{CHP}&= P_{c}+P_{EB}+P_{load}, \\
	 	h_{CHP}+ h_{GB} +h_{EB}+h_{d}&=h_{load}+h_{c}\,.
	 \end{align}
 	\label{eq: pow-balance}
	 \end{subequations}
At any given time $t$, the exact electrical generation $P_{PV}, P_{WT}$ and power demands $P_{load}$ and $h_{load}$ are assumed to be unknown and given by their forecast. 

\subsection{Modeling the electricity market}
\begin{figure}[ht!]
	\centering
	\includegraphics[width=0.8\linewidth]{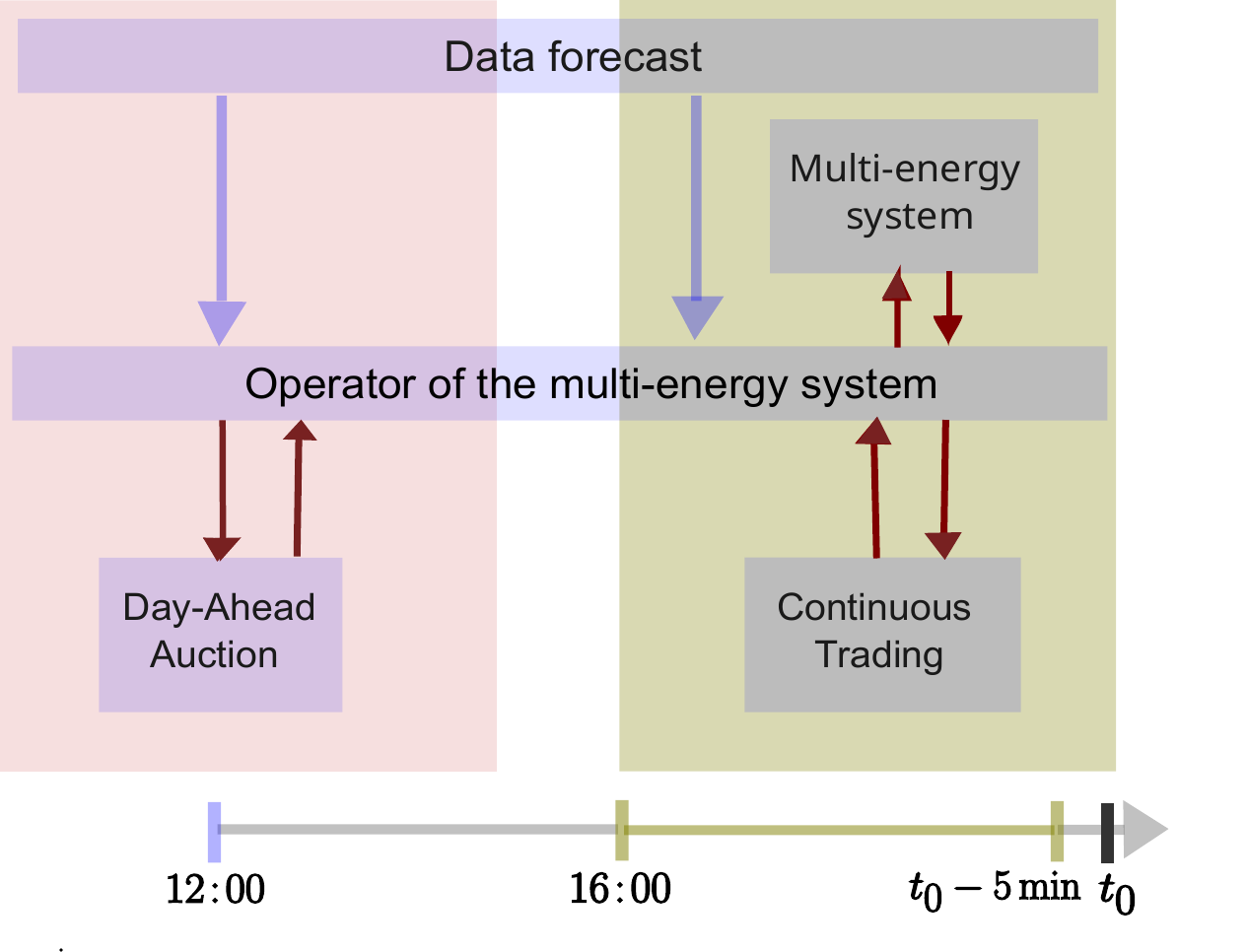}%
	\caption{Overview of the electricity markets for the European market and German market area. Here $t_0$ is the time of fulfillment.}
	\label{fig: electricity-market}
\end{figure}

In this work, we study the sequential electricity market for the European market and German market area, where the electrical grid power is first purchased in the DA market based on an estimate of the electrical/heat generation, demand and DA market prices. In a second step, CT performs corrections with respect to the purchased DA electrical power based on more accurate estimates of the unknown data. A summary of the electricity market under study is depicted in Figure~\ref{fig: electricity-market}.

\subsubsection{Day Ahead Market}
On the day before fulfillment at midday, i.e., 12:00, the MMES operator submits an electricity scheduling plan and signs a contract with the DA market. The scheduling plan is established upon available forecasts of the power generation from the renewable resources, namely photovoltaic and wind turbines,  electrical/heat power demand and the apriori unknown DA prices. 
To formulate the DA optimization problem, the following ingredients are required.

\paragraph{Objective function}
We minimize the cost for participating in the DA market and operating ESS, HSS as well as EB, GB and CHP units as follows
\begin{subequations}
\begin{align}
	J^{DA}:=\sum_{t=0}^{N^{DA}-1} &\left(C^{DA}_e( t) +C^{DA}_{ESS}( t)+C^{DA}_{HSS}( t)+C^{DA}_{GAS}( t) \right),\\
	C^{DA}_e( t)&= \bigg(\frac{\hat v^{DA}(t)-\hat u^{DA}(t)}{2} \vert P^{DA}_g( t)\vert \nonumber \\
	&+\frac{\hat v^{DA}(t)+\hat u^{DA}(t)}{2} P^{DA}_g( t)\bigg) \delta^{DA}, \label{eq: monetary-DA} \\
	C^{DA}_{ESS}(t)&= \rho_{ESS}\,(P^{DA}_{c}( t)+ P^{DA}_{d}( t))\delta^{DA}, \\
	C^{DA}_{HSS}(t)&= \rho_{HSS}\,(h^{DA}_{c}( t)+ h^{DA}_{d}( t)) \delta^{DA}, \\
	C^{DA}_{GAS}( t)&= v_{GAS} \left(\frac{P^{DA}_{CHP}( t)}{\eta_{CHP}}+\frac{h^{DA}_{GB}( t)}{\eta_{GB}}\right) \delta^{DA}.
\end{align}
\label{eq: cost-DA}
\end{subequations}%
 Note that the purchasing $v^{DA}(t)\in\real$ and selling $u^{DA}(t)\in\real$ electricity prices at DA are assumed to be unknown and given by the forecasts $\hat v^{DA}(t)\in\real$ and $\hat u^{DA}(t)\in\real$, $N^{DA}$ is the discrete-time prediction horizon and $\delta^{DA}= 1$h the sampling period of the DA market. Furthermore, we denote by $\rho_{ESS},\,\rho_{HSS}>~0$ the depreciation coefficients of the ESS and HSS, respectively and set them to be 0.01\,\euro{}/ kWh. The gas price $v_{GAS}$ is assumed to be known and given by 0.055\,\euro{}/ kWh as in~\cite{WANG2022118528}. The parameters $\eta_{GB},\eta_{CHP}>0$ denote the gas-to-heat efficiency of the gas boiler and the gas-to-electricity efficiency of the CHP, respectively.

It is noteworthy that the formulation of the DA market cost $C_e(t)\in\real$, found also in~\cite{WANG2022118528} and~\cite{9171315}, takes positive or negative values depending on the purchasing/selling prices $\hat v^{DA}(t), \hat u^{DA}(t)$ and the purchased grid power $P^{DA}_g$. The rationale behind this formulation is that, for positive prices, a positive grid power implies that a given purchased power $P^{DA}_g(t)$ costs $C_e(t) = \hat v^{DA}(t) P^{DA}_g(t) \delta^{DA}$. On the other hand, a negative grid power indicates a monetary gain by $C_e(t) = \hat u^{DA}(t) P^{DA}_g(t) \delta^{DA}$ resulting from selling the excess of electrical power and injecting $\vert P^{DA}_g(t)\vert$ back into the grid. The power balances~\eqref{eq: pow-balance} determine if we purchase from or sell electrical power to the DA market.
We remark that the dispatched power values for the DA scheduling are not actually realized and only $P^{DA}_g$ is forwarded to the optimization problem solving for the CT. Furthermore, we include the operation costs of electricity, heat and gas systems in~\eqref{eq: cost-DA} to find an optimal $P^{DA}_g$ that also keeps the costs of operating the multi-energy system minimal. Minimizing the {market} cost $C_e(t)$ solely as done in~\cite{WANG2022118528} does not achieve this compromise. Finally, observe that in the objective function $J^{DA}$, there are no tracking but only economic, i.e., system and market objectives, where the purchased grid power $P^{DA}_g$ is determined to satisfy the power balance~\eqref{eq: pow-balance}. 

\paragraph{Optimization problem} 
To formulate the optimization problem of the DA market, we collect the decision variables in the following vectors
\begin{align*}
	\mathbf{x}^{DA}&= \big[ 
		(x^{DA}_{SoC})^\top,\, (H^{DA})^\top
		\big]^\top,\\
	\mathbf{u}^{DA} &= \big[(P^{DA}_{c})^\top,(P^{DA}_{d})^\top, (\alpha^{DA})^\top,(h^{DA}_{c})^\top,\\& (h^{DA}_{d})^\top, (\beta^{DA})^\top, (h^{DA}_{EB})^\top, (h^{DA}_{GB})^\top, (P^{DA}_g)^\top, (P^{DA}_{CHP})^\top\big]^\top, 
\end{align*}
as well as the DA forecasts of the unknown generation, demand and DA market prices, 
\begin{align*}
	\mathbf{w}^{DA} = \big[ (\hat P^{DA}_{load})^\top,(\hat P^{DA}_{PV})^\top,\,(\hat P^{DA}_{WT})^\top, (\hat h^{DA}_{load})^\top, (\hat v^{DA})^\top, (\hat u^{DA})^\top \big]^\top.
\end{align*}
The vectors $x^{DA}_{SoC}$ and $H^{DA}$ are of length $N^{DA}+1$ and the vectors in $\mathbf{u}^{DA}$ and $\mathbf{w}^{DA}$ are of length $N^{DA}$. In summary, we arrive at the following optimization problem for the DA scheduling 
\begin{subequations}
\label{eq: day-ahead}
\begin{align}
	&\min_{\mathbf{u}^{DA}} \quad J^{DA}\\
	&\text{s.t.}\quad \mathbf{x}^{DA}\in \mathcal{X}(\mathbf{u}^{DA}), \; \mathbf{u}^{DA}\in \mathcal{U}(\mathbf{w}^{DA}),\\
	& \begin{bmatrix}
		x^{DA}_{SoC}(0),\, H^{DA}(0)
	\end{bmatrix}^\top = \begin{bmatrix}
	x^m_{SoC},\ H^m 
	\label{eq: DA-IC}
	\end{bmatrix}^\top\,.
\end{align}
\end{subequations}
Here, we denote by $\mathcal{X}(\mathbf{u}^{DA})$ the set of state constraints given in~\eqref{eq: const-Pc-Pd} and~\eqref{eq: const-hc-hd} with $\delta =\delta^{DA}$ and $\mathcal{U}(\mathbf{w}^{DA})$ the sets of input constraints satisfying~\eqref{eq: const-Pc-Pd}-\eqref{eq: pow-balance}.
Observe that the state constraints (which include the dynamics \eqref{eq: dyn-soc} and \eqref{eq: dyn-h}) depend on the predicted inputs $\mathbf{u}^{DA}$ and the input constraints depend on the available DA forecasts $\mathbf{w}^{DA}$ via~\eqref{eq: pow-balance}. The constraint~\eqref{eq: DA-IC} fixes the initial condition for the predicted trajectory $\mathbf{x}^{DA}$ to the currently measured system states  $\big[
	x^m_{SoC},\, H^m
\big]^\top$.
\subsubsection{Continuous trading}
The CT opens at 16:00 on the day before the delivery and closes 5 min prior to the fulfillment time $t_0$. Throughout this period, the CT remains available to the MMES operator for possible corrections of the scheduled electrical grid power $P_g^{DA}$. In the sequel, we assume that the CT takes place 15 min before the delivery and therefore the improved CT forecasts of the generation, demand and market prices lie sufficiently close to their real-time values. Without loss of generality, the fulfillment $ t_0$ is assumed to start at midnight, i.e., 24:00. Based on an estimate of the electrical grid power from the DA, the MPC for CT solves for the real-time dispatching and operation of the MMES by correcting the error resulting from over- or under-estimating the electrical grid power. 
To formulate the optimization problem for the CT, we define the objective function below.
\paragraph{Objective function}
Here, we minimize the costs associated with the CT and the real-time dispatching of the multi-energy system given by
\begin{subequations}
	\begin{align}
	J^{CT}:=\sum_{t=0}^{N^{CT}-1}& C_e^{CT}(t)+C^{CT}_{ESS}( t)+C^{CT}_{HSS}( t)+C^{CT}_{GAS}( t),\\
	C_e^{CT}( t)&=\bigg(\frac{\hat v^{CT}(t)-\hat u^{CT}(t)}{2}\vert P^{CT}_g( t)-P_{g,int}^{DA}(t)\vert \nonumber \\
	&+\frac{\hat v^{CT}(t)+\hat u^{CT}(t)}{2}  (P^{CT}_g( t)-P_{g,int}^{DA}(t))\bigg)\delta^{CT}, \label{eq: monetary-CT}\\
	C^{CT}_{ESS}( t)&= \rho_{ESS}\left(P^{CT}_{c}( t)+ P^{CT}_{d}( t)\right) \delta^{CT}, \\
	C^{CT}_{HSS}( t)&= \rho_{HSS} \left(h^{CT}_{c}( t)+ h^{CT}_{d}( t)\right) \delta^{CT}, \\
	C^{CT}_{GAS}( t)&= v_{GAS} \left(\frac{P^{CT}_{CHP}( t)}{\eta_{CHP}}+\frac{h^{CT}_{GB}( t)}{\eta_{GB}}\right) \delta^{CT}.
\end{align}
\label{eq: cost-CT}
\end{subequations}
In~\eqref{eq: cost-CT}, $P^{CT}_{g}(t)$ denotes the actual value of the electrical grid power {and $P_{g,int}^{DA}$ the interpolated sequence of the purchased electrical power $P_{g}^{DA}$ since the sampling periods of DA and CT are different {\tg(compare also Section~\ref{subsec: Pgrid_DA})}.} Moreover, $\hat v^{CT}(t),\, \hat u^{CT}(t)\in\real$ are the forecasts of the CT { purchasing and selling} prices, where the real prices are assumed to be unknown. Let the sampling interval $\delta^{CT}=15\min$ and $N^{CT}>0$ be the discrete-time prediction horizon of the CT. The rationale behind the {electricity market cost} $C^{CT}_e(t)$ can be explained as follows. If $P_{g}^{CT}(t)$ was underestimated in the day-ahead scheduling, namely $P_{g}^{CT}(t)-P_{g,int}^{DA}(t)>0$, then $C^{CT}_e(t) = \hat v^{CT}(t)(P_{g}^{CT}(t)-P_{g,int}^{DA}(t)) \delta^{CT}$ implying that the remaining amount of power is purchased from the grid in the CT scheduling and delivered in real-time accordingly. In case of excess of electrical grid power purchased in the day-ahead market, i.e., $P_{g}^{CT}(t)-P_{g,int}^{DA}(t)<~0$, the electricity market cost amounts to  $C^{CT}_e(t) = \hat u^{CT}(t)(P_{g}^{CT}(t)-P_{g,int}^{DA}(t)) \delta^{CT}$ indicating monetary profit and the electrical grid power returns to the grid.  

\paragraph{Optimization problem}
We define the decision variables of the CT optimization problem as
\begin{align*}
	\mathbf{x}^{CT}&=\big[ (x^{CT}_{SoC})^\top,  (H^{CT})^\top \big]^\top \\
	\mathbf{u}^{CT} &= \big[(P^{CT}_{CHP})^\top, (P^{CT}_{c})^\top, (P^{CT}_{d})^\top, (\alpha^{CT})^\top,\\ & (h^{CT}_{c})^\top, (h^{CT}_{d})^\top, (\beta^{CT})^\top, (h^{CT}_{EB})^\top, (h^{CT}_{GB})^\top, (P^{CT}_g)^\top \big]^\top,
\end{align*}
and the forecast of the generation, demand and CT prices 
\begin{align*}
	\mathbf{w}^{CT}\!\! =\!\! \big[ (\hat P^{CT}_{load})^\top, (\hat P^{CT}_{PV})^\top,\,( \hat P^{CT}_{WT})^\top, ( \hat h^{CT}_{load})^\top, (\hat v^{CT})^\top, ( \hat u^{CT})^\top\big]^\top.
\end{align*}
Here $x^{CT}_{SoC}$ and $H^{CT}$ are column vectors of length $N^{CT}+1$ and the vectors in $\mathbf{u}^{CT}$ and $\mathbf{w}^{CT}$ are of length $N^{CT}$. To find the real-time dispatched values to the multi-energy system, we solve the following optimization problem
\begin{subequations}
\label{eq: continuous-trading}
	\begin{align}
		&\min_{\mathbf{u}^{CT}} \quad J^{CT}\\
		&\text{s.t.}\quad \mathbf{x}^{CT}\in \mathcal{X}(\mathbf{u}^{CT}),\; \mathbf{u}^{CT}\in \mathcal{U}(\mathbf{w}^{CT}) \\
		& \big[
			x^{CT}_{SoC}(0),\, H^{CT}(0)
		\big]^\top = \big[
		x^m_{SoC},\, H^m
	\big]^\top\,. \label{eq: CT-IC} 
	\end{align}
\end{subequations}
Here again, $\mathcal{X}(\mathbf{u}^{CT})$ denotes the explicit dependence of the SoC and the stored heat energy dynamical constraints~\eqref{eq: dyn-soc} and~\eqref{eq: dyn-h} on the predicted inputs $\mathbf{u}^{CT}$, where $\delta =\delta^{CT}$. Moreover, $\mathcal{U}(\mathbf{w}^{CT})$ denotes the dependence of the input constraints on the forecast $\mathbf{w}^{CT}$ by means of~\eqref{eq: pow-balance}. The constraint~\eqref{eq: CT-IC} fixes the initial condition for the predicted trajectory $\mathbf{x}^{CT}$ to the currently measured system states $ [x^m_{SoC},\, H^m]^\top$. The resulting optimization problems~\eqref{eq: day-ahead} and~\eqref{eq: continuous-trading} are mixed-integer linear programs with  non-differentiable cost that can be exactly reformulated into a continuously differentiable function~\cite{9171315}.

\section{Hierarchical MPC}
\label{sec: mpc-sol}
In the following, we propose a hierarchical MPC-based scheme to solve the planning and real-time operation of the MMES in the European market and German market area. %

\subsection{MPC for the day-ahead}
Let $T^{DA}= \delta^{DA} N^{DA}$ be the continuous-time DA prediction horizon.
Given a forecast for the generation, demand and DA prices, $\textbf{w}^{DA}$ over the DA prediction horizon, the current state of charge $x^{m}_{SoC}$ and stored heat energy $H^{m}$, the DA optimization problem~\eqref{eq: day-ahead} is solved once a day, i.e., every $24$h at 12:00 in a receding horizon fashion. The prediction horizon is set to be $T^{DA} = (12\, h+T^{DA}_{pred})$, where $T^{DA}_{pred}$ represents the prediction horizon in hours starting from the time of the delivery $t_0$, i.e., at midnight. We forward only the scheduled grid power $\big(P^{DA}_g(12),\dots, P^{DA}_g(12+T_{pred}^{DA}-1)\big)$, i.e., starting from the next day, to the optimization layer for CT after suitable interpolation. %
Optimizing over the first 12 hours in the MPC for DA scheduling  improves the state and input predictions over $T^{DA}$, because the already signed DA market contract on the previous day is based on older measurements and forecasts. %
After the closure of the DA market, 
the incurring monetary costs from the DA planning are determined as follows
\begin{align}
	\label{eq: index-DA}
	\mathcal{I}^{DA} &:=  \sum_{t=12}^{35} \bigg(\frac{v^{DA}(t)-u^{DA}(t)}{2} \vert P^{DA}_{g}( t)\vert \nonumber \\ &+ \frac{v^{DA}(t)+u^{DA}(t)}{2} P^{DA}_{g}(t)\bigg) \delta^{DA}.
\end{align}
Note that $\mathcal{I}^{DA}$ as defined in~\eqref{eq: index-DA} corresponds to the grid power that is actually purchased on the DA market for $24$h.

\subsection{MPC for the continuous trading}
Let $T^{CT}= \delta^{CT} N^{CT}$ denote the continuous-time prediction horizon of CT.
The CT optimization problem~\eqref{eq: continuous-trading} is solved every $\delta^{CT}=15 \min$ and starts 15 min before the real delivery, i.e., a quarter before midnight. Hence, it is given by $T^{CT} = 15 \min + T^{CT}_{pred}$, where $T^{CT}_{pred}$ is the CT prediction horizon in minutes starting from midnight. Given the current measurements of the state of charge $x^{m}_{SoC}$ and stored heat energy $H^{m}$ and available forecasts of the generation, demand and prices over the CT prediction horizon, the MPC optimization problem predicts the input to be applied to~\eqref{eq: dyn-soc} and~\eqref{eq: dyn-h} at the next time instance, i.e., in $15\min$, in particular $\big(P^{CT}_c(1), P^{CT}_d(1), h^{CT}_c(1), h^{CT}_d(1)\big)$. Hence at each iteration, the first input to ESS and HSS in the sequence of decision variables, in particular, $\big(P^{CT}_c(0), P^{CT}_d(0), h^{CT}_c(0), h^{CT}_d(0)\big)$ is fixed to the optimal one-step (15 min) ahead solution of the optimization problem at the previous sampling instance. To find the optimal inputs to the multi-energy system, the CT compensates, whenever needed, for an over- or under-estimation of the grid power with respect to the purchased DA market $P_g^{DA}$ so that an additional power is purchased or an excess power is sold. Once the MPC algorithm for CT has been solved and dispatched to the MMES, we determine the CT performance index by evaluating the CT cost at the real purchasing $v^{CT}\in\real$ and selling prices $u^{CT}\in\real$ 
	\begin{align}
		\mathcal{I}^{CT}&:= \bigg(\frac{v^{CT}(1)-u^{CT}(1)}{2}(P^{CT}_{g}(1)-P^{DA}_{g,int}(1))\nonumber \\
		&+\frac{v^{CT}(1)+u^{CT}(1)}{2} \big\vert P^{CT}_{g}(1)-P^{DA}_{g,int}(1)\big\vert\bigg)\delta^{CT} \nonumber \\
		&+ C^{CT}_{ESS}(1) + C^{CT}_{HSS}(1)+ C^{CT}_{Gas}(1).
		\label{eq: index_CT}
	\end{align}

\section{Case study}
We apply the proposed hierarchical MPC scheme in Section~\ref{sec: mpc-sol} to the example of an MMES from 2019 in the city of Hanover, Germany. Our numerical values are taken from~\cite{WANG2022118528} after adaption to our setting and summarized in Table~\ref{tab: ess-hss-val}.
\begin{table}[ht!]
	\caption{physical parameters of the case study}
	
	\begin{center}
		\resizebox{.9\linewidth}{!}{
			\begin{tabular}{|l|l | l|}
				\hline
				Definition&  Value& Unit \\
				\hline
				\hline
				Capacity of ESS, Q & 2000  & kWh
				\\
				\hline 
				Charging and discharging efficiency of ESS, $\eta_c/\eta_d$ & 0.95  & p.u. \\
				\hline
				Min./max. SoC bounds, $\underline x/ \overline x$ &0.2/0.8 & p.u. \\
				\hline
				Max. charging and discharging power, $\overline P_{ESS}$ & 0/500 & kW \\
				\hline
				\hline
				Charging and discharging efficiency of HSS, $\eta^h_c/\eta^h_d$ & 0.9  &  p.u. \\
				\hline
				Min./max. HSS bounds,  $\underline H/\overline H$ & 100/ 3000  & kWh    \\
				\hline
				Max. charging and discharging power, $\overline h_{HSS}$ & 750& kW\\
				\hline
				\hline
				Heat-to-electricity ratio of CHP, $b$ &1.2 & p.u.  \\
				\hline
				Min./max. output power of CHP,  $\underline P_{CHP}/\overline P_{CHP}$ & 0/1200 & kW \\
				\hline
				Electricity-to-heat efficiency of EB,  $\eta_{EB}$ & 0.95& p.u.\\
				\hline
				Gas-to-heat efficiency of GB,  $\eta_{GB}$ & 0.8& p.u.\\
				\hline
				Gas-to-electricity efficiency of CHP,  $\eta_{CHP}$ &0.4 & p.u.\\
				\hline
				Min./max.  heat power of EB, $\underline h_{EB}/\overline h_{EB}$ & 0/500 & kW  \\
				\hline
				Min./max.  heat power of GB, $\underline h_{GB}/\overline h_{GB}$ & 0/500 & kW \\
				\hline
				\hline
				Min./max. grid power,  $\underline P_g/\overline P_g $ &-2500/2500 &kW \\
				\hline
		\end{tabular}}
	\end{center}
	\label{tab: ess-hss-val}
	\vspace{-10mm}
\end{table} 

\subsection{Real-time measurements}
\label{subsec: real-time-m}

The load and generation profiles are simulated in an hourly grid with the tools from~\cite{ninjarenew},~\cite{npro} and rescaled to the nominal values in~\cite{WANG2022118528}. They are depicted in Figure 3. 
The nominal values correspond to the rated capacities of electrical and heat load, photovoltaic
and wind turbines. These are given by 1300~kW, 1500~kW, 400~kW and 600~kW, respectively. The quarter-hourly powers are linearly interpolated between the hourly values. The hourly data of the market prices are taken from~\cite{smard}. The quarter-hourly CT prices are constantly interpolated between the sampling times.

\subsection{ Electricity market prices}%
 For the day-ahead, we adopt the setting from~\cite{WANG2022118528}, where the forecasted and real selling prices are given by $\hat u^{DA}= 0.9 \,\hat v^{DA}$ and  $u^{DA}= 0.9\, v^{DA}$. This choice encourages the MMES operator to keep the purchased DA electrical power for future times and not selling it right away. 
For the CT, the selling prices are chosen to be $\hat u^{CT}= 0.8 \,\hat u^{DA}$ and $u^{CT}= 0.8 \, u^{DA}$, whereas the purchasing prices follow $\hat v^{CT}= 1.2 \,\hat v^{DA}$ and $ v^{CT}= 1.2 \,v^{DA}$. Intuitively, it is more expensive to purchase the electrical power and less profitable to sell it in the CT than in the DA market. This motivates an improved power scheduling in the DA market. 
\subsection{Inter- and extrapolation of $P_g^{DA}$}
\label{subsec: Pgrid_DA} %
 Once the electrical grid power has been determined in the DA scheduling with sampling period of $1$h, it is forwarded to CT with a sampling period of $15\min$. Therefore an interpolation of $P_g^{DA}$ is required. For this, we keep the hourly DA value constant $\delta^{DA}/\delta^{CT}$ times. %
 For $T_{pred}^{DA} = 48$h, we utilize the second day predictions, i.e., $(P_g^{DA}(36),...,P_g^{DA}(59))$ to find the interpolated CT sequence over the CT prediction horizon that goes beyond the current day. For $T_{pred}^{DA} = 24$h, an extrapolation of $P_g^{DA}$ towards the end of the CT prediction horizon is performed by repeating the last value of the $P_g^{DA}$ sequence.%

\subsection{Initial conditions}
For the DA optimization, the initial SoC and stored heat energy are taken from~\cite{WANG2022118528} and given by $[x^{DA}_{SoC, init}, H^{DA}_{init}] = [0.3, 600]$. To ensure feasibility at time $t=0$, we set $[x^{CT}_{SoC, init}, H^{CT}_{init}]$ as an interpolated value of the predicted states from the DA, one hour before the delivery and the initial states of the MMES to be the predicted trajectories from the first CT iteration.%
\begin{figure}[ht!]
	\centering
	\includegraphics[width =\linewidth]{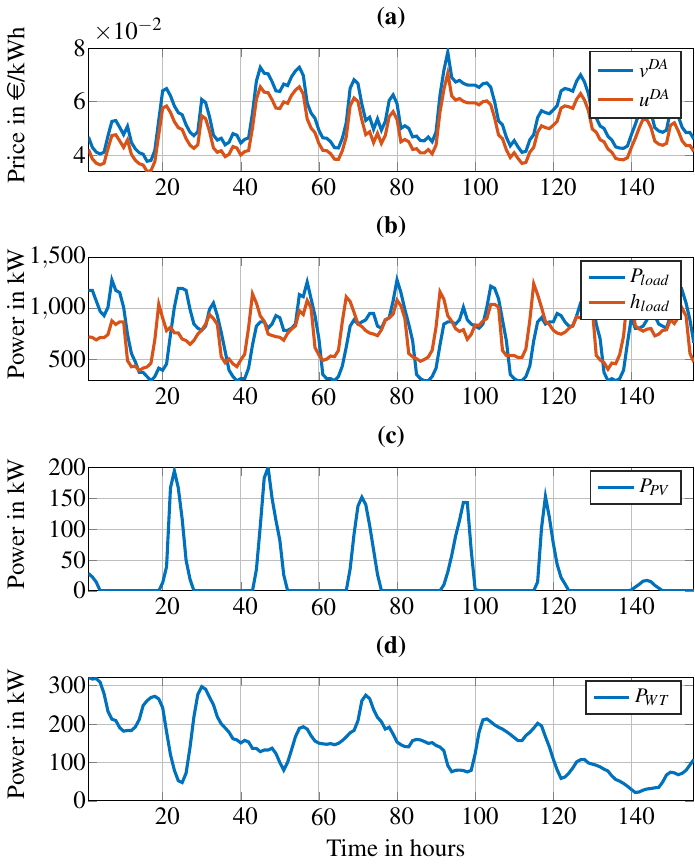}
	\caption{Real measurement data of the (a) purchasing and selling prices, (b) electrical and heat power {loads}, (c) photovoltaic and (d) wind turbine power generation~\cite{ninjarenew},~\cite{npro},~\cite{smard}.}
	\label{fig: real_data}
\end{figure}
\section{Simulation results}
  We consider different horizon lengths for the DA, \mbox{$T_{pred}^{DA}\in\{24\text{h},\, 48\text{h}\}$}. For the CT, we choose \mbox{$T_{pred}^{CT}\in\{1\text{h},\,24\text{h}\}$}. The total simulation time is set to be {4} days. For the evaluation indices in~\eqref{eq: index-DA} and~\eqref{eq: index_CT}, the first day is discarded to limit the effect of the chosen initial conditions. 
 Table~\ref{tab: all} summarizes the simulation results. Therein, the overall performance indices $\mathcal{I}^{DA}_{TOT}, \mathcal{I}^{CT}_{TOT}$ are obtained by evaluating the DA and CT costs~\eqref{eq: index-DA},~\eqref{eq: index_CT}, respectively and building the sum over the total simulation time. %
 
\subsection{Perfect forecast}
We first illustrate the proposed hierarchical MPC scheme for the particular choice of perfect forecast, i.e., where the forecasted data corresponds to the real values for both DA market and CT. The results are presented in Table~\ref{tab: all}. 
\begin{table}[ht!]
	
	\caption{Simulation results of the MMES under study over 4 days}
	\centering Noiseless case 
	\begin{center}
		\resizebox{0.65\linewidth}{!}{
			\begin{tabular}{|l |l || l| l|}
				\hline
				($T_{pred}^{DA}$, $T_{pred}^{CT}$) &  $\mathcal{I}_{TOT}^{DA}$  & $\mathcal{I}_{TOT}^{CT}$ & $\mathcal{I}_{TOT}^{DA}+\mathcal{I}_{TOT}^{CT}$\\
				\hline
				\hline
				(24h, 1h)  & 1918 &  3856 &  5774
				\\
				\hline
				(24h,  24h)   & 1917   & 3844 &  5761  \\
				\hline
				(48h, 24h)   & 1883 &   3828 & 5711\\
				\hline
		\end{tabular}}
	\end{center}
	\centering Noisy case 1
	\begin{center}
		\resizebox{0.65\linewidth}{!}{
			\begin{tabular}{| l|| l| l | l |}
				\hline
				($T_{pred}^{DA}$ , $T_{pred}^{CT}$) &  $\mathcal{I}_{TOT}^{DA}$  & $\mathcal{I}_{TOT}^{CT}$ & $\mathcal{I}_{TOT}^{DA}+\mathcal{I}_{TOT}^{CT}$\\
				\hline
				\hline
				(24h,  24h)  & 1834  &3943  & 5777 
				\\
				\hline
				(48h, 24h)   &1708 & 4012   &  5720
				\\
				\hline
		\end{tabular}}
	\end{center}
	\centering Noisy case 2
	\begin{center}
		\resizebox{0.65\linewidth}{!}{
			\begin{tabular}{|l || l| l| l |}
				\hline
				($T_{pred}^{DA}$ , $T_{pred}^{CT}$) &  $\mathcal{I}_{TOT}^{DA}$  & $\mathcal{I}_{TOT}^{CT}$ & $\mathcal{I}_{TOT}^{DA}+\mathcal{I}_{TOT}^{CT}$\\
				\hline
				\hline
				(24h,  24h)   &  1757 & 4045   &  5803
				\\
				\hline
				(48h, 24h)   &  1738    & 3993  & 5731   \\
				\hline
		\end{tabular}}
	\end{center}
	\label{tab: all}
	\vspace{-5mm}
\end{table}

\begin{figure}[ht!]
		\centering
		\includegraphics[width =\linewidth]{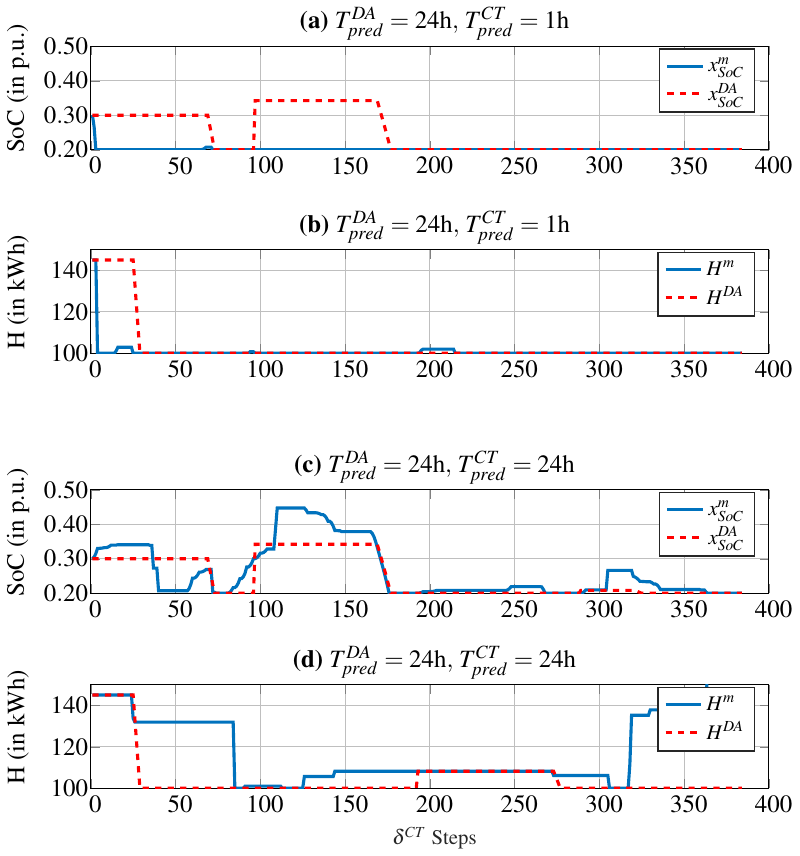}
		\caption{Evolution of the state of charge $x^{DA}_{SoC}$ (a),\,(c) and the stored heat energy $h^{DA}$ (b),(d) in the DA (in red) and in closed-loop (in blue) for $(T_{pred}^{DA}, T_{pred}^{CT})=(24\text{h},1\text{h})$ in {(a),\,(b)} and $(T_{pred}^{DA}, T_{pred}^{CT})=(24\text{h},24\text{h})$ in {(c),\,(d)} over 4 days with perfect forecast.}
		\label{fig: 24-1-24-24}
	\end{figure}

\begin{figure} 
		\centering
		\includegraphics[width =\linewidth]{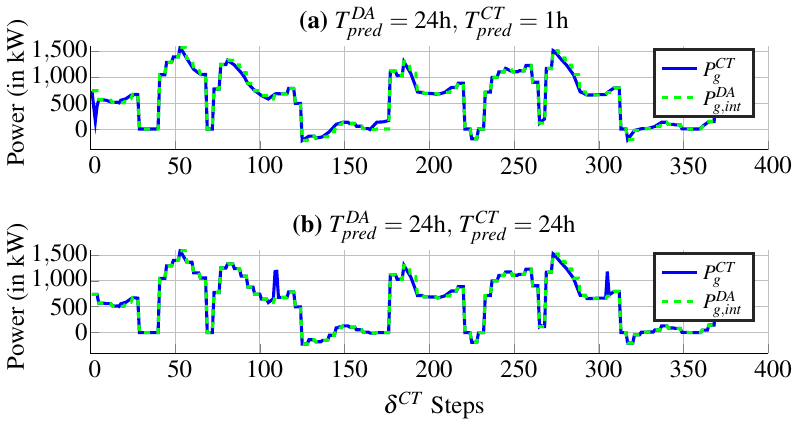}
		\caption{Predicted values of the purchased grid power in the DA (green) and CT (blue) for $(T_{pred}^{DA}, T_{pred}^{CT})=(24\text{h},1\text{h})$ in (a) and $(T_{pred}^{DA}, T_{pred}^{CT})=(24\text{h},24\text{h})$ in (b) over 4 days with a perfect forecast.}
		\label{fig: p_grid}
	\end{figure}

For the same prediction horizon of the DA, e.g., 24h, an increase in the prediction horizon of the CT towards that of the DA, e.g., from 1h to 24h leads to a decrease in DA and CT evaluation indices and therefore their overall sum. This can be explained as follows: a long CT prediction horizon (of the same length as for the DA problem) results in CT predictions that are more closely in line with the DA predictions. On the other hand, using a too short CT prediction horizon can result in a too short-sighted optimization in the CT layer, resulting in dispatched grid powers that are suboptimal in the long run and hence incur a higher overall cost. To illustrate this, Fig.~\ref{fig: 24-1-24-24} compares the state of charge and
stored heat of 
\begin{itemize}
	\item[(ii)] the closed-loop values of the hierarchical MPC scheme in Section~\ref{sec: mpc-sol}, {blue in Fig.~\ref{fig: 24-1-24-24}}
	\item[(i)] {the interpolated trajectory resulting from solving for the following day in the DA market~\eqref{eq: day-ahead}, red in Fig.~\ref{fig: 24-1-24-24}}
\end{itemize}
	for the cases $(T_{pred}^{DA}, T_{pred}^{CT})=(24\text{h},1\text{h})$ and $(T_{pred}^{DA}, T_{pred}^{CT})=(24\text{h},24\text{h})$. Even with perfect forecast, the SoC and heat energy trajectories resulting from the MPC solving the DA scheduling are further away from those of the closed-loop for $T_{pred}^{DA}\gg T_{pred}^{CT} $. { In general, the shortsightedness of CT does not seem to incentivize the usage of the electrical storage system}. {For $T_{pred}^{DA} =  T_{pred}^{CT}$, the closed-loop trajectories remains overall closer to those predicted by the DA scheduling.} For the same prediction horizon in DA and CT, the closed-loop trajectories are not expected to follow exactly the predicted DA trajectories, since we interpolate the real-time power profiles linearly and extrapolate $P_g^{DA}$ towards the end of the CT prediction horizon (see Section~\ref{subsec: Pgrid_DA}). Moreover, the selling/purchasing price ratios are different from DA market to CT. Finally, the input sequences $\mathbf{u}^{CT}$ of the CT have more degrees of freedom than being constantly interpolated between sampling intervals of the $\mathbf{u}^{DA}$ sequences. %
	However, in both settings of the prediction horizons, the purchased grid power from DA remains close to that of the CT as shown in Fig.~\ref{fig: p_grid}. Therein, the visible differences between DA and CT predictions can again be inferred from the reasons named above.
	On the other hand, observe that from Table~\ref{tab: all}, an increase in the DA prediction horizon, e.g., from $24$h to $48$h, leads to a decrease in the DA and CT indices. For the horizon lengths $(48\text{h},24\text{h})$, the overall performance index is the smallest, which thus is a reasonable choice for the prediction horizons.

\subsection{Imperfect forecast}
We now turn to the more realistic case of imperfect data forecast as specified below.

\paragraph{Forecast method}
Since no forecast model for the case study is available, the forecasted data is determined by adding a relatively growing noise level to the real-time power measurements and market prices. The generated noise sequences are drawn from a uniform distribution. 
For load and generation profiles of the electrical and heat systems, the random noise is propagated from 0\% at the beginning of the prediction horizon and linearly increased with the following relatively growing rates after 24h:
\begin{itemize}
	\item Noisy case 1:  10\% 
	\item Noisy case 2: 20\% 
\end{itemize}

We select a noise level from 0\% at the beginning of the prediction horizon to 5\% after $24$h for the purchasing and selling prices. The forecast data across generation and demand within the DA are uncorrelated but correlated to CT forecast data. This choice is motivated by more unpredictable fluctuations in the power generation/demand due to the unknown photovoltaic and wind power and consumption compared to less varying prices regulated by the European market. To obtain the CT forecast, we add a relatively growing noise sequence (according to the noise levels in case 1 and 2) to the CT real-time data over the CT prediction horizon, after linearly interpolating the same random noise sequences used in the DA forecast.

\paragraph{Discussion}

Table~\ref{tab: all} summarizes the simulation results of noisy case~1 and~2. Here again, a too short CT compared to the DA prediction horizon, e.g., $(T_{pred}^{DA},T_{pred}^{CT})=(24\text{h}, 1\text{h})$ is not a suitable choice. In particular, this combination of DA/CT prediction horizons leads to infeasibility of the MPC for CT problem for the noisy case 1 and 2 due to the shortsightedness of the MPC solving for CT. %
In the noisy case 1, at the time of delivery the power forecasts for DA market worsen by 5\% and for the CT by 0.1\% relatively to their real values. In the noisy case 2, they worsen by 10\% and of the CT by 0.21\% relatively to their real (noiseless) values. This results in a higher overall sum of the CT and DA indices and sheds light into the effect of the uncertainty, i.e., demand, generation and market prices forecast on the overall incurring costs across DA and CT.

\section{Conclusion} %
We designed a hierarchical MPC scheme for the predictive operation of multi-energy systems for the European market and German market area. We suggest a suitable choice of the prediction horizon and analyze the impact of imperfect forecasts on the total incurring cost on an energy network located in Hanover, Germany. Future work aims at relaxing our modeling assumptions of nearly perfect knowledge of the real-time data of the CT by adding a third layer for real-time dispatching. Another future venue is to improve the forecast method and take the uncertainty explicitly into account via robust MPC schemes.

\bibliographystyle{ieeetran.bst}
\bibliography{root.bib}

@Article{JI2021107133,
  author  = {Yongli Ji and Qingshan Xu and Jun Zhao and Yongbiao Yang and Lu Sun},
  title   = {Day-ahead and intra-day optimization for energy and reserve scheduling under wind uncertainty and generation outages},
  journal = {Electric Power Systems Research},
  year    = {2021},
  volume  = {195},
  pages   = {107133},
  issn    = {0378-7796},
  doi     = {https://doi.org/10.1016/j.epsr.2021.107133},
}

@Misc{ninjarenew,
  author = {Stefan Pfenninger and Iain Staffell},
  title  = {{Renewables.ninja}},
  year   = {2023},
  note   = {\url{https://www.renewables.ninja/}},
}

@Book{rawlings2017model,
  author    = {Rawlings, J.B. and Mayne, D.Q. and Diehl, M.},
  title     = {Model Predictive Control: Theory, Computation, and Design, 2nd Edition},
  year      = {2017},
  publisher = {Nob Hill Publishing},
  isbn      = {9780975937730},
}

@Article{SCHULZE2016592,
  author  = {Tim Schulze and Ken McKinnon},
  title   = {The value of stochastic programming in day-ahead and intra-day generation unit commitment},
  journal = {Energy},
  year    = {2016},
  volume  = {101},
  pages   = {592-605},
  issn    = {0360-5442},
  doi     = {https://doi.org/10.1016/j.energy.2016.01.090},
}

@Article{LINGMIN2020118536,
  author  = {Chen Lingmin and Wu Jiekang and Wu Fan and Tang Huiling and Li Changjie and Xiong Yan},
  title   = {Energy flow optimization method for multi-energy system oriented to combined cooling, heating and power},
  journal = {Energy},
  year    = {2020},
  volume  = {211},
  pages   = {118536},
  issn    = {0360-5442},
  doi     = {https://doi.org/10.1016/j.energy.2020.118536},
  url     = {https://www.sciencedirect.com/science/article/pii/S0360544220316443},
}

@Article{8231216,
  author  = {Vasilj, Josip and Gros, Sebastien and Jakus, Damir and Zanon, Mario},
  title   = {Day-Ahead Scheduling and Real-Time Economic MPC of CHP Unit in Microgrid With Smart Buildings},
  journal = {IEEE Transactions on Smart Grid},
  year    = {2019},
  volume  = {10},
  number  = {2},
  pages   = {1992-2001},
  doi     = {10.1109/TSG.2017.2785500},
}

@Article{9171315,
  author  = {Garcia-Torres, Felix and Bordons, Carlos and Tobajas, Javier and Marquez, Juan Jose and Garrido-Zafra, Joaquin and Moreno-Munoz, Antonio},
  title   = {Optimal Schedule for Networked Microgrids Under Deregulated Power Market Environment Using Model Predictive Control},
  journal = {IEEE Transactions on Smart Grid},
  year    = {2021},
  volume  = {12},
  number  = {1},
  pages   = {182-191},
  doi     = {10.1109/TSG.2020.3018023},
}

@Misc{npro,
  title = {{nPro}: Planning tool for buildings \& districts},
  year  = {2023},
  note  = {\url{https://app.npro.energy/}},
}

@Misc{smard,
  title = {{SMARD}: {S}trommarktdaten},
  year  = {2023},
  note  = {\url{https://www.smard.de}},
}

@Article{WANG2022118528,
  author   = {Yubin Wang and Wei Dong and Qiang Yang},
  title    = {Multi-stage optimal energy management of multi-energy microgrid in deregulated electricity markets},
  journal  = {Applied Energy},
  year     = {2022},
  volume   = {310},
  pages    = {118528},
  issn     = {0306-2619},
  doi      = {https://doi.org/10.1016/j.apenergy.2022.118528},
}

\end{document}